\documentclass[10pt,superscriptaddress,reprint,amsmath,amssymb,aps,prl,showpacs,onecolumn]{revtex4-2}
\usepackage{mathrsfs}
\usepackage{graphicx}
\usepackage[caption=false,farskip=0pt,labelfont={bf}]{subfig}
\usepackage{dcolumn}
\usepackage{bm}
\usepackage{amssymb}
\usepackage{amsmath}
\usepackage{paralist}
\usepackage{amsmath,amssymb,mathrsfs}

\usepackage[colorlinks,linkcolor=blue,anchorcolor=blue,citecolor=green]{hyperref}
\usepackage{cleveref}
\usepackage{url}
\usepackage{color}

\makeatletter

\renewcommand{\maketag@@@}[1]{\hbox{\m@th\normalsize\normalfont#1}}%

\makeatother

\begin{document}

\title{Exact Spectral Function of One-Dimensional Bose Gases}

\author{Song Cheng}
\affiliation{Department of Physics, The University of Hong Kong, Hong Kong, China}
\affiliation{Beijing Computational Science Research Center, Beijing 100193, China }

\author{Yang-Yang Chen}
\email[]{chenyy@nwu.edu.cn}
\affiliation{Shaanxi Key Laboratory for Theoretical Physics Frontiers,  Xi'an 710069, China}
\affiliation{Institute of Modern Physics, Northwest University, Xi'an 710069, China}

\author{Xi-Wen Guan}
\email[]{xiwen.guan@anu.edu.au}
\affiliation{Innovation Academy for Precision Measurement Science and Technology, Chinese Academy of Sciences, Wuhan 430071, China}
\affiliation{Department of Fundamental and Theoretical Physics, Research School of Physics, Australian National University, Canberra ACT 0200, Australia}
\affiliation{Peng Huanwu Center for Fundamental Theory, Xi'an 710069, China}

\author{Wen-Li Yang}
\affiliation{Shaanxi Key Laboratory for Theoretical Physics Frontiers,  Xi'an 710069, China}
\affiliation{Institute of Modern Physics, Northwest University, Xi'an 710069, China}
\affiliation{Peng Huanwu Center for Fundamental Theory, Xi'an 710069, China}

\author{Rubem Mondaini}
\affiliation{Department of Physics, University of Houston, Houston, Texas}

\author{Hai-Qing Lin}
\email[]{haiqing0@csrc.ac.cn}

\affiliation{School of Physics, Zhejiang University, Hangzhou 310058, China}

\date{\today}


\begin{abstract}
Exactly solved models provide rigorous understanding of many-body phenomena in  strongly correlated systems.   In this article, we report a breakthrough in uncovering  universal many-body correlated properties of quantum integrable Lieb-Liniger model. We calculate exactly the dynamical correlation functions  by computing  the form factor through a newly developed method, by which  we are capable of calculating all possible ‘relative excitations’ over the ground state or a  finite temperature state at a high precision. Consequently, full spectral functions obtained for the model manifests the unique power-law singularity behaviour at the spectral threshold, confirming the validity of nonlinear Luttinger liquid theory.  Our method  advances the theory of dynamical correlation functions with high precision towards the thermodynamic limit, and is capable of benchmarking experimental observation of such novel correlated properties [1].
\end{abstract}

\maketitle

\noindent

The novel phenomena associated with strongly correlated systems such as Mott phase transition, spin-charge separation and Fermi edge singularity (FES) have fascinated numerous effort of condensed matter physicists \cite{Mahan,Gogolin}.
A significant way to enhance the correlation for interacting quantum systems is to reduce the spatial dimensionality, which simplifies complicated situations in reality and thereby helps a lot in resolving the puzzles \cite{Giuliani,Giamarchi}.
In this regard,  the key  to understanding the  underlying physics  of  quantum many-body systems  is to  discover behaviour of correlation functions.
Nevertheless, from  a theoretical perspective, despite very few limiting cases such as Tonks-Girardeau gas with various confinements \cite{Girardeau,Lenard,Tracy1979,Pezer,Minguzzi}, rigorous  calculation of correlation functions remains a formidable task, even on an account of the  results obtained by quantum Monte Carlo (QMC) and density matrix renormalization group (DMRG).
From an experimental perspective, in  the past few decades, one has witnessed successful developments of the ultracold-atoms with unprecedented levels of manipulation and control \cite{Bloch,Guan:2022}.
So far only a few kinds  of correlation functions in one-dimensional (1D) systems have been measured, such as the momentum distribution via optical imaging \cite{Paredes2004,Yang:2017}, spectral function via photoemission spectroscopy \cite{ps} and momentum-resolved Raman spectroscopy \cite{Dao}, two-point correlation function of dynamical structure factor (DSF) via Bragg scattering spectroscopy \cite{Bs,Fabbri2015,Senaratne:2021,Meinert:2015} etc.
Such novel progress has therefore stimulated a great demand on exact results for miscellaneous correlation functions in 1D systems, also see recent observation of the dissipative dynamics in 1D interacting ultracold atoms \cite{Zhao-Tsinghua:2023}.

Based on the Bethe ansatz  \cite{QISM,Guan2013}, recent developments of the non-perturbative physics of exactly solved models either in or out of equilibrium are very stimulating \cite{CA2016,Doyon2018,Nardis2018,Bertini2021,Doyon2023,Nardis2023}.
 However, despite decades of extensive research, a rigorous understanding and exact computation of dynamical correlation functions at a many-body level   have remained elusive.
In this paper, by extending  the {\em form factor} to large-size systems  and  dividing  the infinite  dimensions of the Hilbert space into a series of subspaces  for arbitrary  {\em `relative excitations',  a highly efficient algorithm is established for   calculating exactly  many-body correlated  properties of quantum integrable systems.
As an exotic application, we compute rigorously  various correlation functions of the 1D Bose gases,  i.e. the Lieb-Liniger model, at unprecedented  system sizes, } giving  exact results of the spectral function beyond the accuracy of the so far existing numerical calculations of  correlation functions for this model.
Our algorithm is as powerful as successfully capturing the power-law singularities in the vicinities of spectral thresholds, and showing  certainly different advantages from other methods, for example, the ABACUS developed by Caux and his collaborators \cite{Caux2006,Caux2007,Panfil,Caux2009}.
Our results suggest that only a system size up to thousands of particles can guarantee enough resolution for reaching the true power-law of the correlations at the spectral edges, thus further confirming the validity of the nonlinear Tomonaga-Luttinger liquid \cite{Imambekov2008,Imambekov2009,Imambekov2009b,Imambekov2012}, and ensuring a rigorous access to the emergent behavior of correlation functions \cite{Li2023} in the thermodynamic limit.

\section{MODEL}\label{sec_model}
We now apply our algorithm to calculate the spectral function of the Lieb-Liniger model (LLM) \cite{Lieb} that  describes $N$ bosons confined on a line of length $L$ with contact interaction.
The exact solution of the model \cite{QISM,Jiang} benchmarks a large variety of many-body phenomena, ranging from universal thermodynamics and quantum criticality \cite{Jiang} to correlation functions \cite{QISM,Slavnov1997,Kojima1997,Korepin1984,Slavnov,Caux2006,Panfil,Li2023,Caux2007,SC,Granet:2021,Gora2003a,Cheianov:2006,Nadani:2016,Kormos:2009}, and the behavior of Tomonaga-Luttinger liquid (TLL) \cite{Imambekov2008,Imambekov2009b,Imambekov2012,Kitanine}.
Such theoretical developments have inspired unprecedented levels of experimental study in the ultracold atoms
\cite{Yang:2017,Fabbri2015,Meinert:2015}, see recent reviews \cite{Guan:2022}.

The Hamiltonian of LLM reads
\begin{equation}\label{hamiltonian}
H = - \sum_{i=1}^{N} \frac{\partial^2}{\partial x_i^2} + 2c \sum_{i>j}^{N} \delta \left( x_i - x_j \right)
\end{equation}
where $c>0$ ($c<0$) stands for repulsion (attraction), and a dimensionless parameter $\gamma=c*L/N$  is introduced for the interaction strength \cite{Lieb}.
Hereafter only the repulsive interaction will be taken into account.
With the help of the Bethe ansatz, solving the eigenvalue problem of the Hamiltonian (\ref{hamiltonian}) reduces to solving the transcendental Bethe ansatz equations (BAEs)
\begin{equation}\label{BAE}
\lambda_j + \frac{1}{L} \sum_{k=1}^{N} \theta \left( \lambda_j - \lambda_k \right) = \frac{2\pi}{L} I_j, \quad j = 1, \dots, N
\end{equation}
where $\theta(x)=2\arctan(x/c)$, pseudomomenta $\{\lambda_j\}$ are distinct real numbers and QNs $\{I_j\}$ are distinct integers (half-integers) if N is odd (even).
There is a one-to-one map between a set of QNs and a set of pseudomomenta, by utilizing which the total momentum and energy of system are expressed as
$P_{\{\lambda\}}=\sum_{j=1}^{N} \lambda_j, \, E_{\{\lambda\}}=\sum_{j=1}^{N} \lambda_j^2$.
The ground state is formulated by a Fermi sea-like distribution for QNs (i.e. $I=\{ -\frac{N-1}{2},\dots, \frac{N-1}{2} \}$).
In this model, $P_\mathrm{m}$, one of the tag quantum numbers, connects the excited momentum through $P_\mathrm{m}=P*L/2\pi$.

The spectral function in general represents the probability of tunneling a particle with specified momentum and energy into or out of the system.
Let us start from the single particle Green's function
\begin{equation}\label{def_GF}
\textmd{i}\cdot\mathcal{G}(x,t) \equiv \langle \mathcal{T} \left[ \Psi(x,t)\Psi^\dagger(0,0) \right] \rangle_N
\end{equation}
where $\langle \cdots \rangle_N$ means expectation value taken over the ground state of $N$-particle system, and $\Psi(x,t)$ is the bosonic field operator.
For simplicity, we merely consider the larger Green's function $G^>(x,t)$, and the treatment for the lesser one is similar.
Inserting a completeness relation into the two field operators gives rise to
$\textmd{i}\cdot G^> (x,t) = \sum_{\{\mu\}_{N+1}} \frac{ \langle \{ \lambda \}_{N} | \Psi(x,t) |  \{ \mu\}_{N+1}\rangle \,   \langle \{ \mu \}_{N+1} | \Psi^\dagger(0,0) |  \{ \lambda\}_N\rangle }{ \langle \{ \lambda \}_N| \{ \lambda\}_N\rangle \, \langle \{ \mu \}_{N+1}| \{ \mu\}_{N+1}\rangle}$
where $| \{ \nu \}_M \rangle$ is an eigenstate consisting of $M$ particles and specified by a set of pseudomomenta $\{ \nu\}_M$.
The ground state and intermediate state are denoted respectively as $| \{ \lambda \}_N \rangle$ and $| \{ \mu\}_{N+1} \rangle$.
The {\em form factor} of field operator is $\mathcal{F}(\{\lambda\}_N,\{\mu\}_{N+1})= \langle \{ \mu \}_{N+1} | \Psi^\dagger(0,0) |  \{ \lambda\}_N\rangle$.
Based on these notations, we have
\begin{eqnarray}
\textmd{i} \cdot G^>(x,t) = \sum_{\{\mu\}_{N+1}} e^{\textmd{i} \phi^+ } \frac{|\mathcal{F}(\{\lambda\}_N,\{\mu\}_{N+1})|^2}{\| \{ \mu\}_{N+1} \|^2 \cdot \| \{\lambda\}_N \|^2}
\end{eqnarray}
with $\phi^+= (E_{\{\lambda\}}-E_{\{\mu\}})t-(P_{\{\lambda\}}-P_{\{\mu\}})x$.
According to the definition of SF $A(k,\omega) = -\frac{1}{\pi} \textmd{Im} \mathcal{G}(k,\omega)$, where $\mathcal{G}(k,\omega)$ is the Fourier transform of $\mathcal{G}(x,t)$, one finally obtains
\begin{footnotesize}
\begin{align}\label{A}
\frac{A(k,\omega)}{L} =  \sum_{\{\mu\}_{N+1}}  \frac{  \delta_{k,P_{ \{\mu \},\{\lambda\}}}    \delta(\omega - E_{\{\mu\},\{\lambda\}} )    |\mathcal{F}(\{\lambda\}_N,\{\mu\}_{N+1})|^2}{\| \{ \mu\}_{N+1} \|^2 \cdot \| \{\lambda\}_N \|^2} \notag\\
+
 \sum_{\{\mu\}_{N-1}}  \frac{  \delta_{-k,P_{ \{\mu \},\{\lambda\}}}    \delta(\omega +  E_{\{\mu\},\{\lambda\}} )    |\mathcal{F}(\{\mu\}_{N-1},\{\lambda\}_{N})|^2}{\| \{ \mu\}_{N-1} \|^2 \cdot \| \{\lambda\}_N \|^2}
\end{align}
\end{footnotesize}
where $\delta_{n,m}$ is the Kronecker delta function and $C_{\{\mu\},\{ \lambda \}}\equiv C_{\{ \mu \}}-C_{\{\lambda\}}$ if $C=P$ or $E$.
The validity of the result is quantitatively checked by the saturation of sum-rule
\begin{equation}\label{sum_rule}
\sum_{k} \int_{-\infty}^{0}  \frac{\textmd{d}\omega}{ N} A(k,\omega) = 1.
\end{equation}

\section{ALGORITHM}\label{sec_algo}
Given an operator $\hat{O}$, its {\em form factor} is defined by $\mathcal{F} (|r\rangle,|s\rangle) = \langle s | \hat{O} |r \rangle$, i.e. the matrix element evaluated between two eigenstates of the interacting Hamiltonian \cite{QISM}.
The time-dependent correlation function is  expressed as following spectral representation
\begin{footnotesize}
\begin{equation}\label{spectral_representation}
\langle s| \hat{O}^\dagger(x,t) \hat{O}(0,0) | s \rangle = \sum_{|r\rangle \in \mathfrak{E}} e^{\mathrm{i} (E_{r,s}t - P_{r,s}x)} \frac{|\mathcal{F}(|r\rangle, |s\rangle) |^2} { \| s \|^2 \, \| r \|^2}
\end{equation}
\end{footnotesize}
where $|s\rangle$ is the eigenstate under study, $\mathfrak{E}$ is the eigenspace of the Hamiltonian, $E_{r,s}=E_r-E_s$  ($P_{r,s}=P_r -P_s$) stands for the difference of energy (momentum) between $|r\rangle$ and $|s\rangle$, and $\| \dots \|$ is the norm of a state.
The state $|s\rangle$ is not limited to the ground state, while in some circumstances such as finite temperature and post-quench, it may be a highly excited state.

Note that the summation is assumed to include all the eigenstates, and the evaluation of Eq.~\ref{spectral_representation} is nothing but counting the elements of $\mathfrak{E}$  together with calculating their form factors.
It is obvious that the key step  is to efficiently and quickly find the essential states in the process of navigating $\mathfrak{E}$. To this end, we have developed an algorithm suitable for
calculating various dynamical correlation functions, such as the dynamical structure factor in ground state \cite{Li2023} and the one-body dynamical correlation at finite temperatures \cite{SC}.
The seminal idea is that the most relevant states for calculating form factors of a local observable ought not to be much different from the state of our interest in the perspective of QNs configuration.
In light of the prospectively unified description for any state under investigation which later on is called {\em reference state}, one has to abandon the conventional understanding of the particle-hole pairs of excitation over the Fermi sea.
We introduce the concept of `relative excitation' that leads to the re-distribution of QNs away from the reference state, see Supplementary Materials for an explicit example.

For the purpose of classifying the excited states over a reference state, we introduce a set of four tags ($P_m, N_p, P_l, N_l$). They are four non-negative integers, the wave number $P_m = k L/2\pi$ specifies the excited momentum, $N_p$ is the number of particles involved in the `relative excitation', $N_l < N_p$ is the number of particles jumping leftward, and $P_l \geq N_l$ is the sum of excited momentum due to those $N_l$ leftward particles in units of
$2\pi/L$.
They can be seen as a set of QNs to describe the `relative excitation’ over a reference state. In this way, $\mathfrak{E}$ is separated into a large number of subspaces, and one may find the most relevant states through a proper choice of tags.
%
%
Since this partition for $\mathfrak{E}$ dynamically depends on the reference state, our algorithm is very efficient in accelerating the search of those relevant states that make non-negligible contributions, see SM for the example Table.
Consequently, this method is especially efficient for tackling the highly excited states, for instance at finite temperature or the post-quench steady state.
Moreover, our algorithm considers momentum as the first quantum number, which makes itself a convenient access to the line-shape of dynamical correlations as well as directly benchmarking the experimental data.
All these features distinct from ABACUS \cite{Caux2009} manifest the capacity of our algorithm through following natural observations on the power-law behavior of the spectral function at the spectral edge.


\section{RESULTS}\label{sec_results}

\begin{figure*}[htbp]
\includegraphics[width=0.45\linewidth]{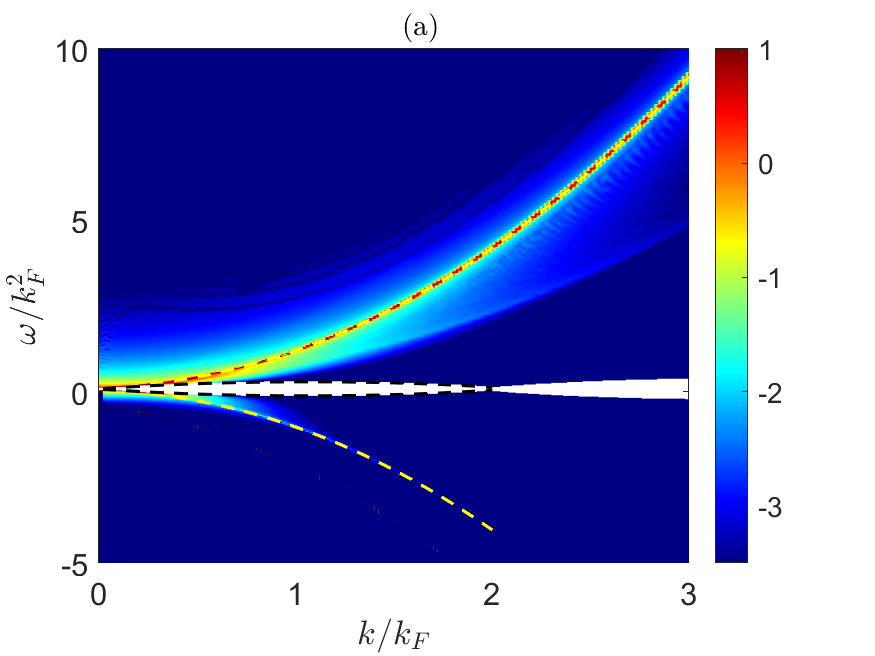}
\includegraphics[width=0.45\linewidth]{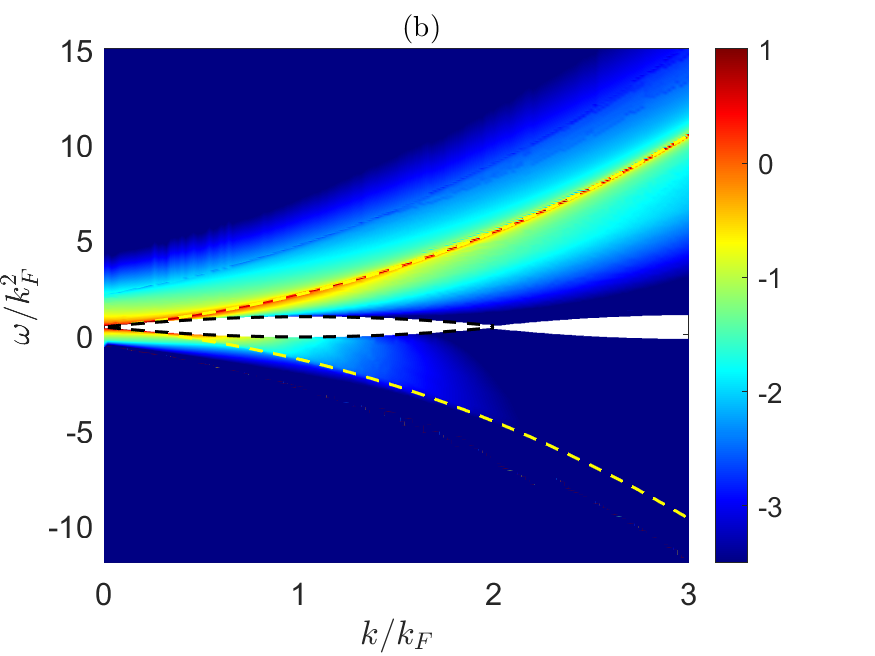}
\caption{The momentum-energy resolved SF of Lieb-Liniger gas in ground state of system size  $N=L=100$. The subfigures (a) and (b) are for interaction strength $\gamma=0.5$ and $4.0$ respectively with sum rules $0.9999$ and $0.9935$.
Momentum and energy are given  in units  of Fermi energy and Fermi momentum,  respectively. For the sake of clarity,  we adopt the logarithm of SF, showing the higher the value the brighter the color.
The yellow (black) dashed lines are the Type-I (-II) dispersion relations.}
\label{contour_plot}
\end{figure*}

\begin{figure*}[htbp]
\includegraphics[width=0.5\linewidth]{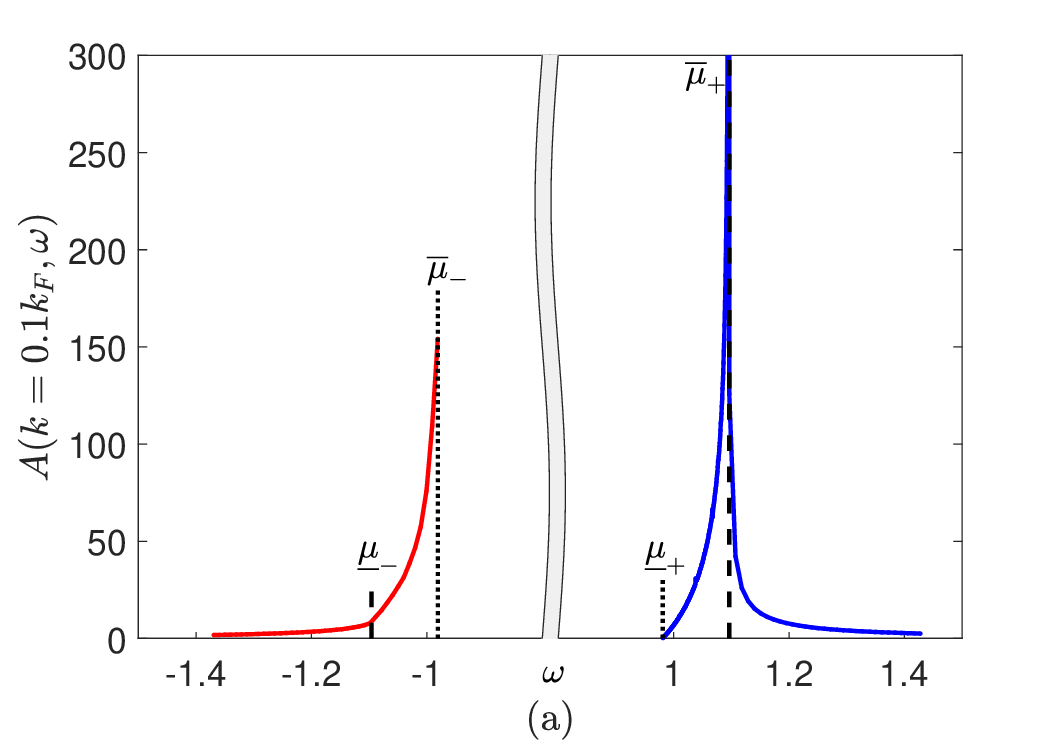}
\includegraphics[width=0.5\linewidth]{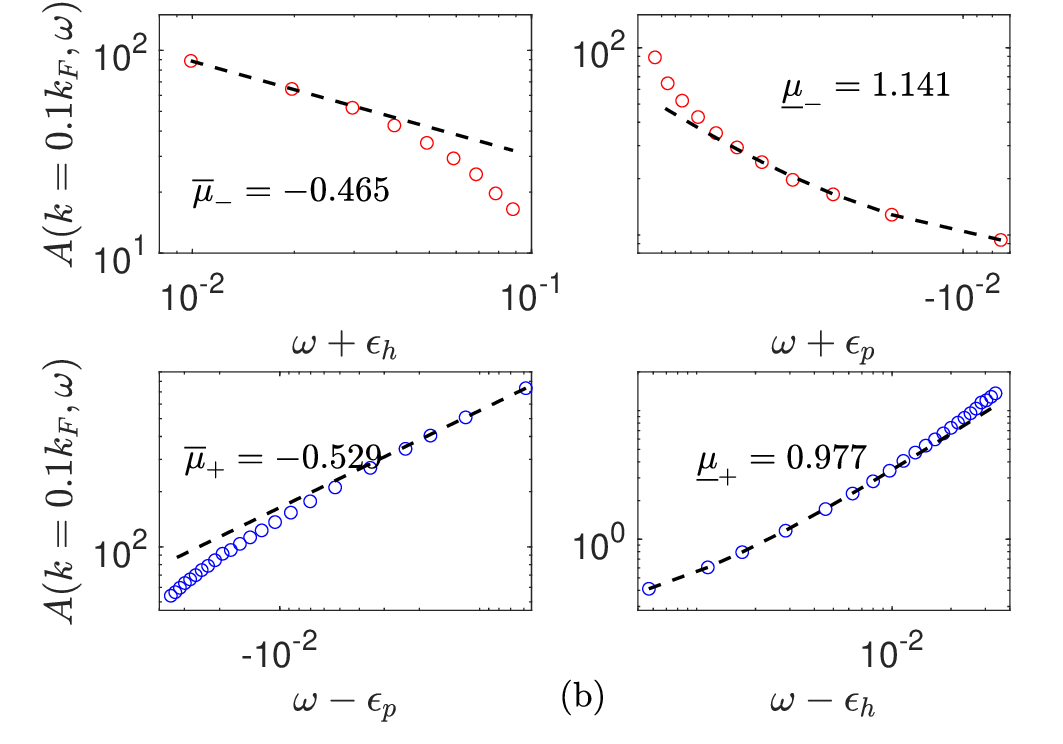}
\caption{SF vs energy for the momentum  $k=0.1k_F$ and the  interaction strength $\gamma=4.0$. The system size is $N=L=4000$, and the energy is measured in units of Fermi energy.
(a) shows the full power-law feature  of the  SF  with  varying the energy, where red (blue) curve stands for the emission (absorption) spectrum, and black dashed (dotted) lines are the thresholds of single particle spectrum $\pm \epsilon_p$ ($\pm \epsilon_h$).
Apparently, there exists a peak around Type-I (-II) dispersion in absorption (emission) spectrum.
Between the two Type-II excitations  is blank indicated in Fig.~\ref{contour_plot}.
(b) shows the  singularity powers of SF in the vicinities of  the thresholds of single-particle spectrum.  A log-log coordinate is used for a sake of clear visibility. The red (blue) circles represent SF in emission (absorption) spectrum and black dashed lines are power-law curves predicted by Eq.~\ref{FES}, where the corresponding exponents $\overline{\mu}_-$, $\underline{\mu}_-$, $\overline{\mu}_+$, and $\underline{\mu}_+$ are $-0.465$, $1.141$, $-0.529$, and $0.977$, in good agreement with the prediction of nonlinear TLL  $-0.422$, $0.934$, $-0.501$, and $1.043$, respectively.
}
\label{c4Line_Power}
\end{figure*}

The logarithm of  SF for  the LLM is demonstrated in Fig.~\ref{contour_plot} with the system size $N=L=100$. Here the focus is concentrated on the intermediate interaction and  weak interaction situations i.e. $\gamma=4.0$ and $\gamma=0.5$ respectively.
The yellow  (black) dashed lines stand for the Type-I (-II) dispersion relations, corresponding to the creation of a particle (hole)  outside (inside) of the Fermi sea \cite{Lieb,QISM}.
Single-particle and -hole excitations defining the edges of spectra feature the non-trivial role of interaction in 1D many-body systems.
%
In the comparison of Fig.~\ref{contour_plot}(a) and \ref{contour_plot}(b), the energy of hole excitation is suppressed when the interaction strength $\gamma$  decreases,  but the Type-II dispersion exists as long as $\gamma$ does not vanish.
This phenomenon underlies why the Bogoliubov approximation is deficient and the explicit explanation was first put forward by the Bethe ansatz solution for this model \cite{Lieb}.
It is obvious that either absorption or emission spectrum is separated into three regions all of which are determined by pairs of p-h excitations.
Let us denote the Type-I (-II) dispersion as $\epsilon_p$ ($\epsilon_h$) in the particle sector $\omega >0$, and as $-\epsilon_p$ ($-\epsilon_h$) in the hole sector $\omega<0$.
For the half plane of $\omega>0$, below $\epsilon_h$ is blank, implying no excitation bears that energy and momentum which is seen clearly in Fig.~\ref{c4Line_Power} as well.
For $\epsilon_h < \omega < \epsilon_p$ a spectral continuum consisting of states produced by arbitrary pairs of p-h excitations occurs, while for $\omega>\epsilon_p$, it is the region where the excitations involve the states generated by $2$-, $3$-, and multiple p-h pairs.

Shown in Fig.~\ref{c4Line_Power}(a) is the line-shape of SF vs energy with fixed momentum $k=0.1k_F$ and interaction strength $\gamma=4.0$.
It should be noted that the system size here is $N=L=4000$, different from Fig.~\ref{contour_plot}.
The particle-hole asymmetry is evidenced by the blue (red) curve within absorption (emission) spectrum performing a peak around the Type-II (-I) dispersion.
The mechanism of their birth is distinct that the red peak mainly comes from the single p-h excitation while the blue from the situation of multiple p-h pairs.

In Fig.~\ref{c4Line_Power}(b) the fascinating many-body phenomenon FES is clearly observed, which is a typical impurity problem closely related to orthogonal catastrophe \cite{Mahan,Giamarchi}.  The FES manifests  that the SF on the thresholds of spectra displays power-law behavior \cite{Imambekov2008,Imambekov2009b,Imambekov2012}
\begin{equation}\label{FES}
A(k,\omega) \sim \textmd{const} + \left| \omega \mp \epsilon_{p,h} \right| ^{\mu_\pm}
\end{equation}
where subscript $+$ (-) specifies the edge exponent on absorption (emission) threshold.
The original FES arises from the transient potential brought forth by a deep electron excitation, which leaves behind a core hole and scatters with the non-interacting electrons in the conduction band \cite{Mahan}.
It was interpreted as an impurity problem, i.e.  an impurity moving in a Fermi liquid. By the inclusion of interaction between particles, one reformed it into a Luttinger liquid instead \cite{Gogolin}.
However, the conventional TLL only identifies the power law, and  its particle-hole symmetry at the long wave length limit  prevents itself from distinguishing four thresholds \cite{Giamarchi,Gogolin}.
This ambiguity disappears if we take account of the nonlinearity of spectrum by combining the Bosonization and quantum integrable theory together \cite{Imambekov2008}.
In our method, it is obvious that the threshold behavior of SF in the thermodynamic limit naturally emerges only when the energetic resolution narrows down to and even lower than $10^{-2} k_F^2$, which in turn requires a very large system size $N=L=4000$.
Using the log-log coordinates, the exponents represented by dashed lines are readily obtained: $\overline{\mu}_-$, $\underline{\mu}_-$, $\overline{\mu}_+$, and $\underline{\mu}_+$ are $-0.465$, $1.141$, $-0.529$, and $0.977$.
Our results for first time confirm the validity of the power-law behavior predicted from the nonlinear TLL with the exponents, explicitly $-0.422$, $0.934$, $-0.501$, and $1.043$ \cite{Imambekov2008}.
This reveals that the nonlinear TLL only validates in such a tiny region.
In contrast, our algorithm is highly capable of evaluating the dynamical correlations with high precision in the whole energy region.

\begin{figure*}[!htbp]
  \center
  \includegraphics[width=1.0\linewidth]{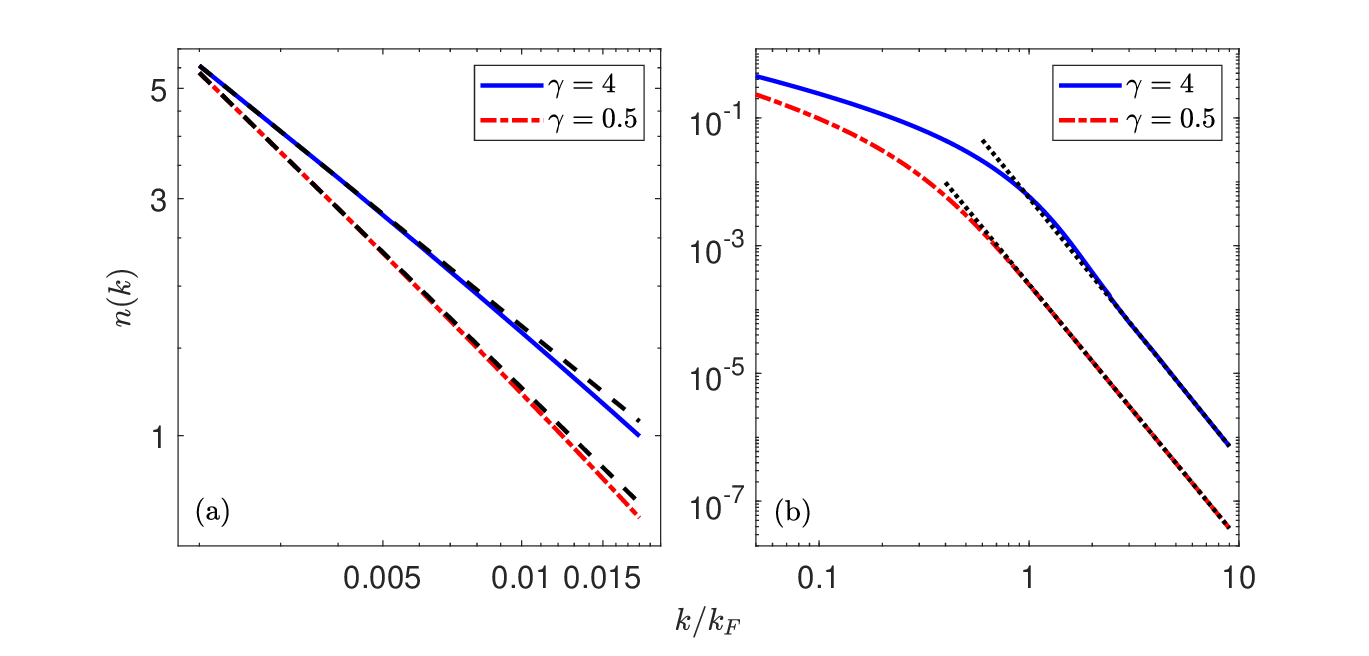}
  \caption{(a) The MDF in the TLL  region. The black dashed lines represent the asymptotic powers  $n(k) \sim k^\alpha $  with $\alpha =-0.750$ and $-0.908$ for $\gamma=4.0$ and $0.5$,  respectively. Here the momentum is very small  $k \rightarrow 0$. The numerical result of the powers  further confirm  the TLL prediction  $\alpha =0.737$ and $0.894$. In this figure  we  set   $N=L=1000$ for our numerical calculation.
 (b) The  MDF in the large momentum regime. The gradients of black dotted lines show  the asymptotic powers for a  large momentum, $\alpha =-4.073$ and $-4.001$ for $\gamma=4.0$ and $0.5$,  respectively. They  agree well with power-law for the large  momentum tail $k^{-4}$.
 The Tan's contact is extracted as well $\mathcal{C}=0.595$ and $0.024$ for $\gamma=4.0$ and $0.5$ respectively, in good agreement with the theoretical prediction $0.524$ and $0.024$.
 Here the system size is  set as $N=L=100$.}
  \label{momentum_distribution}
\end{figure*}

Moreover, in Fig.~\ref{momentum_distribution},  we study the power-law behavior of momentum distribution function (MDF), i.e. the static correlation $n(k)=\int_{-\infty}^{0}  \frac{ \text{d}\omega }{2\pi L}  A(k,\omega)$  in the small and large momentum limits.
The tail of MDF for a system of contact interaction fulfills the universal law, Tan's relation $\lim \limits_{k \rightarrow \infty} n(k) = \mathcal{C} k^{-4}$. Here the weight $\mathcal{C}$ is the Tan's contact which builds wide relations with the quantities such as the internal energy, pair correlation function and pressure.
This asymptotic power-law behavior is shown in Fig.~\ref{momentum_distribution} (b), with the same system size $N=L=100$ as Fig.~\ref{contour_plot}.
The exponents are extracted by the black dotted lines, with the gradients $-4.00$ and $-4.07$ for $\gamma=0.5$ and $4.0$ respectively.
The Tan's contact $\mathcal{C}$ is theoretically calculated through thermodynamic Bethe ansatz equations, and $\mathcal{C}=0.524$ and $0.024$ for $\gamma=4$ and $0.5$ respectively, in accordance with $0.595$ and $0.024$ obtained from Fig.~\ref{momentum_distribution} (b).
For the small momentum region shown in Fig.~\ref{momentum_distribution} (a), it is well known that the MDF obeys an asymptotic power-law that $\lim \limits_{k \rightarrow 0} n(k) \sim k^{\frac{1}{2K}-1}$ according to the TLL theory \cite{Giamarchi}, and easy calculation gives the powers $-0.737$ and $-0.894$.
Since the data of Fig.~\ref{contour_plot} are not sufficient for a visible resolution in momentum, we therefore make use of a larger system $N=L=1000$.
The gradients of black dashed lines are $-0.750$ and $-0.908$ for $\gamma=4.0$ and $\gamma=0.5$ respectively.
Here we show that the results of different methods agree well with each other.
This again indicates the high capability of our method to rigorously study the emergent behavior of correlation functions appearing in the thermodynamic limit.
A wide application of our finding to other integrable systems within the form factor formalism is straightforward.

\section{CONCLUSIONS}\label{sec_conclusion}

The evaluation of correlation functions for a strongly correlated system in general is challenging.
In this scenario, the quantum integrable models are of significant importance to benchmark the observations in ultracold atomic experiments and other solid state materials.
In this paper, building on the {\em form factors} and Bethe ansatz solution, we have presented an efficient algorithm to rigorously calculate the correlated properties of quantum integrable systems.  In particular, we have obtained the spectral function of the Lieb-Liniger Bose gases with arbitrarily interaction strength in the so far arguably the best precision.
The spectral distribution on the full momentum-energy plane, the line-shapes, and especially the power-law behavior of the dynamic correlations on spectral thresholds are explicitly presented.
The power law of the static correlation function, alias the MDF in the large and small momentum region is obtained, as well as the Tan's contact $\mathcal{C}$, showing an excellent agreement with the theoretical prediction.

It is worth emphasizing that the exponents describing Fermi edge singularities have been given explicitly for the system with the largest accessible size $N=L=4000$, which essentially confirms the nonlinear TLL.
We have observed that such a huge system is indeed capable of necessitating the power-law of correlation functions, so far beyond the capability of other methods.
Hundreds of particles cannot guarantee the validation of such a power-law.

Our algorithm efficiently deals with the Fermi-sea like quantum numbers, and is widely applicable to  other quantum integrable systems with nested structure such as the 1D SU(2) Fermi gases.
Moreover, combining with the local density approximation (LDA), it is well suited for systems with confinements, such as Bose gases in harmonic and linear traps.

In general, our work provides a rigorous approach to various emergent features of correlation functions for 1D strongly correlated systems either in or out of equilibrium such as spectral function, dynamical structure factor and quench dynamics, etc. This also opens new avenues for experimental verification in quantum gases and condensed matter systems.

\section*{ACKNOWLEDGMENTS}
SC is grateful to Andrea Trombettoni and Feng He for helpful discussions.

\section*{FUNDING}
This work is supported by the National Natural Science Foundation of China Grants No. 92365202, No. 12134015, No. 12104372, No. 12088101, No. 12047511, No. 12247103; HK GRF Grants No. 17306024 and No. 17313122, CRF Grants No. C4050-23GF and No. C7012-21G, and RGC Fellowship Award No. HKU RFS2223-7S03.

\section*{AUTHOR CONTRIBUTIONS}
S.C. and Y.Y.C. conceived the project, and did the theoretical and numerical calculations.
All authors contributed to the discussion and finalization of the manuscript.

\noindent \textit{\textbf{Conflict of interest statement}}. None declared.


\begin{thebibliography}{10}

\bibitem{Zhao-Tsinghua:2023}
Zhao Y, Tian Y, Ye J {\em et~al.}
\newblock Universal dissipative dynamics in strongly correlated quantum gases.
\newblock {\em Nat. Phys.}, 2025; 1-6.


\bibitem{Mahan}
Mahan GD.
\newblock {\em Many-Particle Physics}. New York: Springer, 2000.

\bibitem{Gogolin}
Gogolin AO, Nersesyan AA and Tsvelik AM.
\newblock {\em Bosonization and Strongly Correlated Systems}. Cambridge: Cambridge University Press, 1998.

\bibitem{Giuliani}
Giuliani G and Vignale G.
\newblock {\em Quantum Theory of The Electron Liquid}, Cambridge: Cambridge University Press, 2005.

\bibitem{Giamarchi}
Giamarchi T.
\newblock {\em Quantum Physics in One Dimension}, Oxford: Oxford University Press, 2004.

\bibitem{Girardeau}
Girardeau M.
\newblock Relationship between systems of impenetrable bosons and fermions in
  one dimension.
\newblock {\em J. Math. Phys.}  1960; {\bf 1}: 516--523.

\bibitem{Lenard}
Lenard A.
\newblock Momentum distribution in the ground state of the one-dimensional
  system of impenetrable bosons.
\newblock {\em J. Math. Phys.}  1964; {\bf 5}: 930--943.

\bibitem{Tracy1979}
Vaidya HG and Tracy CA.
\newblock One-particle reduced density matrix of impenetrable bosons in one
  dimension at zero temperature.
\newblock {\em Phys. Rev. Lett.}  1979; {\bf 42}: 3--6.

\bibitem{Pezer}
Pezer R and Buljan H.
\newblock Momentum distribution dynamics of a Tonks-Girardeau gas: Bragg
  reflections of a quantum many-body wave packet.
\newblock {\em Phys. Rev. Lett.}  2007; {\bf 98}: 240403.

\bibitem{Minguzzi}
Settino J, Gullo NL, Plastina F {\em et~al.}
\newblock Exact spectral function of a Tonks-Girardeau gas in a lattice.
\newblock {\em Phys. Rev. Lett.}  2021; {\bf 126}: 065301.

\bibitem{Bloch}
Bloch I, Dalibard J and Zwerger W.
\newblock Many-body physics with ultracold gases.
\newblock {\em Rev. Mod. Phys.}  2008; {\bf 80}: 885.


\bibitem{Guan:2022}
Guan XW and He P.
\newblock New trends in quantum integrability: recent experiments with
  ultracold atoms.
\newblock {\em Rep. Prog. Phys.}  2022; {\bf 85}: 114001.

\bibitem{Paredes2004}
Paredes B, Widera A, Murg V {\em et~al.}
\newblock Tonks-Girardeau gas of ultracold atoms in an optical lattice.
\newblock {\em Nature}  2004; {\bf 429}: 277--281.

\bibitem{Yang:2017}
Yang B, Chen YY, Zheng YG {\em et~al.}
\newblock Quantum criticality and the tomonaga-luttinger liquid in
  one-dimensional bose gases.
\newblock {\em Phys. Rev. Lett.}  2017; {\bf 119}: 165701.

\bibitem{ps}
Stewart JT, Gaebler JP and Jin DS.
\newblock Using photoemission spectroscopy to probe a strongly interacting
  fermi gas.
\newblock {\em Nature}  2008; {\bf 454}: 744--747.

\bibitem{Dao}
Dao TL, Georges A, Dalibard J {\em et~al.}
\newblock Measuring the one-particle excitations of ultracold fermionic atoms
  by stimulated raman spectroscopy.
\newblock {\em Phys. Rev. Lett.}  2007; {\bf 98}: 240402.

\bibitem{Bs}
Veeravalli G, Kuhnle E, Dyke P {\em et~al.}
\newblock Bragg spectroscopy of a strongly interacting fermi gas.
\newblock {\em Phys. Rev. Lett.}  2008; {\bf 101}: 250403.

\bibitem{Fabbri2015}
Fabbri N, Panfil M, Cl\'{e}ment D {\em et~al.}
\newblock Dynamical structure factor of one-dimensional bose gases:
  Experimental signatures of beyond-luttinger-liquid physics.
\newblock {\em Phys. Rev. A}  2015; {\bf 91}: 043617.

\bibitem{Senaratne:2021}
Senaratne R, Cavazos-Cavazos D, Wang S {\em et~al.}
\newblock Spin-charge separation in a one-dimensional fermi gas with tunable
  interactions.
\newblock {\em Science}  2022; {\bf 376}: 1305--1308.

\bibitem{Meinert:2015}
Meinert F, Panfil M, Mark MJ {\em et~al.}
\newblock Probing the excitations of a lieb-liniger gas from weak to strong
  coupling.
\newblock {\em Phys. Rev. Lett.}  2015; {\bf 115}: 085301.


\bibitem{QISM}
Korepin VE, Bogoliubov NM and Izergin AG.
\newblock {\em Quantum Inverse Scattering Method and Correlation Functions}, Cambridge: Cambridge University Press, 1993.

\bibitem{Guan2013}
Guan XW, Batchelor MT and Lee CH.
\newblock Fermi gases in one dimension: From bethe ansatz to experiments.
\newblock {\em Rev. Mod. Phys.}  2013; {\bf 85}: 1633--1691.

\bibitem{CA2016}
Castro-Alvaredo OA, Doyon B and Yoshimura T.
\newblock Emergent hydrodynamics in integrable quantum systems out of
  equilibrium.
\newblock {\em Phys. Rev. X}  2016; {\bf 6}: 041065.

\bibitem{Doyon2018}
Doyon B, Yoshimura T and Caux JS.
\newblock Soliton gases and generalized hydrodynamics.
\newblock {\em Phys. Rev. Lett.}  2018; {\bf 120}: 045301.

\bibitem{Nardis2018}
Nardis JD, Bernard D and Doyon B.
\newblock Hydrodynamic diffusion in integrable systems.
\newblock {\em Phys. Rev. Lett.}  2018; {\bf 121}: 160603.

\bibitem{Bertini2021}
Bertini B, Heidrich-Meisner F, Karrasch C {\em et~al.}
\newblock Finite-temperature transport in one-dimensional quantum lattice
  models.
\newblock {\em Rev. Mod. Phys.}  2021; {\bf 93}: 025003.

\bibitem{Doyon2023}
Doyon B, Perfetto G, Sasamoto T {\em et~al.}
\newblock Emergence of hydrodynamic spatial long-range correlations in
  nonequilibrium many-body systems.
\newblock {\em Phys. Rev. Lett.}  2023; {\bf 131}: 027101.

\bibitem{Nardis2023}
Nardis JD, Gopalakrishnan S and Vasseur R.
\newblock Nonlinear fluctuating hydrodynamics for Kardar-Parisi-Zhang scaling
  in isotropic spin chains.
\newblock {\em Phys. Rev. Lett.}  2023; {\bf 131}: 197102.

\bibitem{Caux2006}
Caux JS and Calabrese P.
\newblock Dynamical density-density correlations in the one-dimensional bose
  gas.
\newblock {\em Phys. Rev. A}  2006; {\bf 74}: 031605(R).

\bibitem{Caux2007}
Caux JS, Calabrese P and Slavnov NA.
\newblock One-particle dynamical correlations in the one-dimensional bose gas.
\newblock {\em J. Stat. Mech.} 2007; P01008.

\bibitem{Panfil}
Panfil M and Caux JS.
\newblock Finite-temperature correlations in the Lieb-Liniger one-dimensional
  bose gas.
\newblock {\em Phys. Rev. A}  2014; {\bf 89}: 033605.

\bibitem{Caux2009}
Caux JS.
\newblock Correlation functions of integrable models: A description of the
  abacus algorithm.
\newblock {\em J. Math. Phys.}  2009; {\bf 50}: 095214.

\bibitem{Imambekov2008}
Imambekov A and Glazman LI.
\newblock Exact exponents of edge singularities in dynamic correlation
  functions of 1d bose gas.
\newblock {\em Phys. Rev. Lett.}  2008; {\bf 100}: 206805.

\bibitem{Imambekov2009}
Imambekov A and Glazman LI.
\newblock Universal theory of nonlinear luttinger liquids.
\newblock {\em Science}  2009; {\bf 323}: 228--231.

\bibitem{Imambekov2009b}
Imambekov A and Glazman LI.
\newblock Phenomenology of one-dimensional quantum liquids beyond the
  low-energy limit.
\newblock {\em Phys. Rev. Lett.}  2009; {\bf 102}: 126405.

\bibitem{Imambekov2012}
Imambekov A, Schmidt TL and Glazman LI.
\newblock One-dimensional quantum liquids: Beyond the luttinger liquid
  paradigm.
\newblock {\em Rev. Mod. Phys.}  2012; {\bf 84}: 1253--1306.

\bibitem{Li2023}
Li RT, Cheng S, Chen YY {\em et~al.}
\newblock Exact results of dynamical structure factor of Lieb-Liniger model.
\newblock {\em J. Phys. A: Math. Theor.}  2023; {\bf 56}: 335204.

\bibitem{Lieb}
Lieb EH and Liniger W.
\newblock Exact analysis of an interacting bose gas.
\newblock {\em Phys. Rev.}  1963; {\bf 130}: 1605--1615.

\bibitem{Jiang}
Jiang YZ, Chen YY and Guan XW.
\newblock Understanding many-body physics in one dimension from the
  lieb-liniger model.
\newblock {\em Chin. Phys. B}  2015; {\bf 24}: 050311.

\bibitem{Slavnov1997}
Kojima T, Korepin VE and Slavnov NA.
\newblock Determinant representation for dynamical correlation functions of the
  quantum nonlinear schrodinger equation.
\newblock {\em Commun. Math. Phys.}  1997; {\bf 188}: 657--689.

\bibitem{Kojima1997}
Kojima T, Korepin VE and Slavnov NA.
\newblock Completely integrable equation for the quantum correlation function
  of nonlinear schrodinger equation.
\newblock {\em Commun. Math. Phys.}  1997; {\bf 189}: 709--728.

\bibitem{Korepin1984}
Korepin VE.
\newblock Correlation functions of the one-dimensional bose gas in the
  repulsive case.
\newblock {\em Commun. Math. Phys.}  1984; {\bf 94}: 93--113.

\bibitem{Slavnov}
Slavnov NA.
\newblock Calculation of scalar products of wave functions and form factors in
  the framework of the algebraic bethe ansatz.
\newblock {\em Theor. Math. Phys.}  1989; {\bf 79}: 502--508.

\bibitem{SC}
Cheng S, Chen YY, Guan XW {\em et~al.}
\newblock One-body dynamical correlation function of the Lieb-Liniger model at
  finite temperature.
\newblock {\em Phys. Rev. A}  2025; {\bf 111}: L010802.

\bibitem{Granet:2021}
Granet E.
\newblock Low-density limit of dynamical correlations in the Lieb-Liniger
  model.
\newblock {\em J. Phys. A: Math. Theor.}  2021; {\bf 54}: 154001.

\bibitem{Gora2003a}
Kheruntsyan KV, Gangardt DM, Drummond PD {\em et~al.}
\newblock Pair correlations in a finite-temperature 1d bose gas.
\newblock {\em Phys. Rev. Lett.}  2003; {\bf 91}: 040403.

\bibitem{Cheianov:2006}
Cheianov VV, Smith H and Zvonarev MB.
\newblock Three-body local correlation function in the Lieb-Liniger model:
  bosonization approach.
\newblock {\em J. Stat. Mech.} 2006; P08015.

\bibitem{Nadani:2016}
Nandani EKJP, R\"{o}mer R, Tan S {\em et~al.}
\newblock Higher-order local and non-local correlations for 1d strongly
  interacting bose gas.
\newblock {\em New J. Phys.}  2016; {\bf 18}: 055014.

\bibitem{Kormos:2009}
Kormos M, Mussardo G and Trombettoni A.
\newblock Expectation values in the Lieb-Liniger bose gas.
\newblock {\em Phys. Rev. Lett.}  2009; {\bf 103}: 210404.

\bibitem{Kitanine}
Kitanine N, Kozlowski KK, Maillet JM {\em et~al.}
\newblock Form factor approach to dynamical correlation functions in critical
  models.
\newblock {\em J. Stat. Mech.} 2012; P09001.

\end{thebibliography}

\begin{figure*}[ht]
 \centering
 \includegraphics[width=\textwidth]{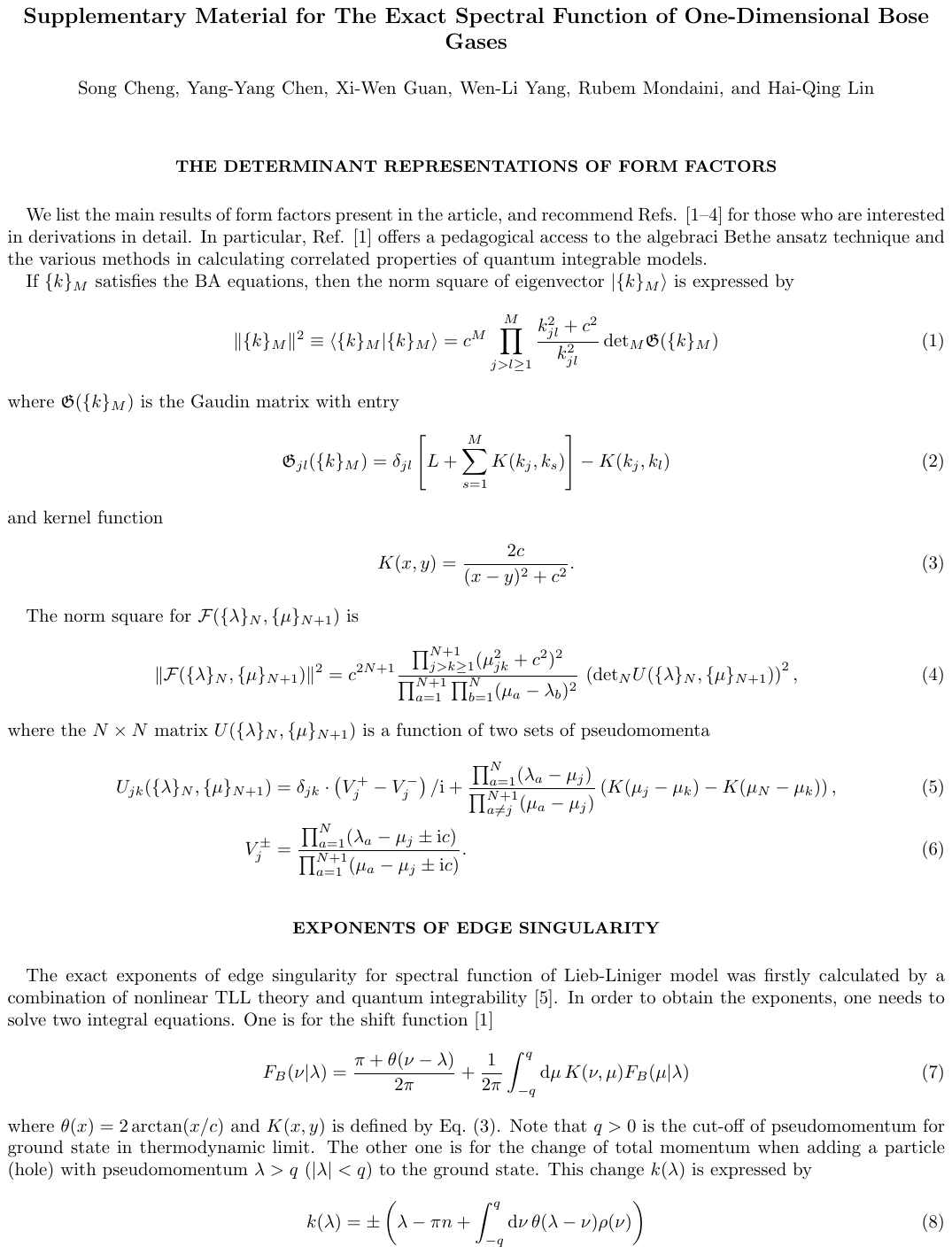}
\end{figure*}

\begin{figure*}[ht]
 \centering
 \includegraphics[width=\textwidth]{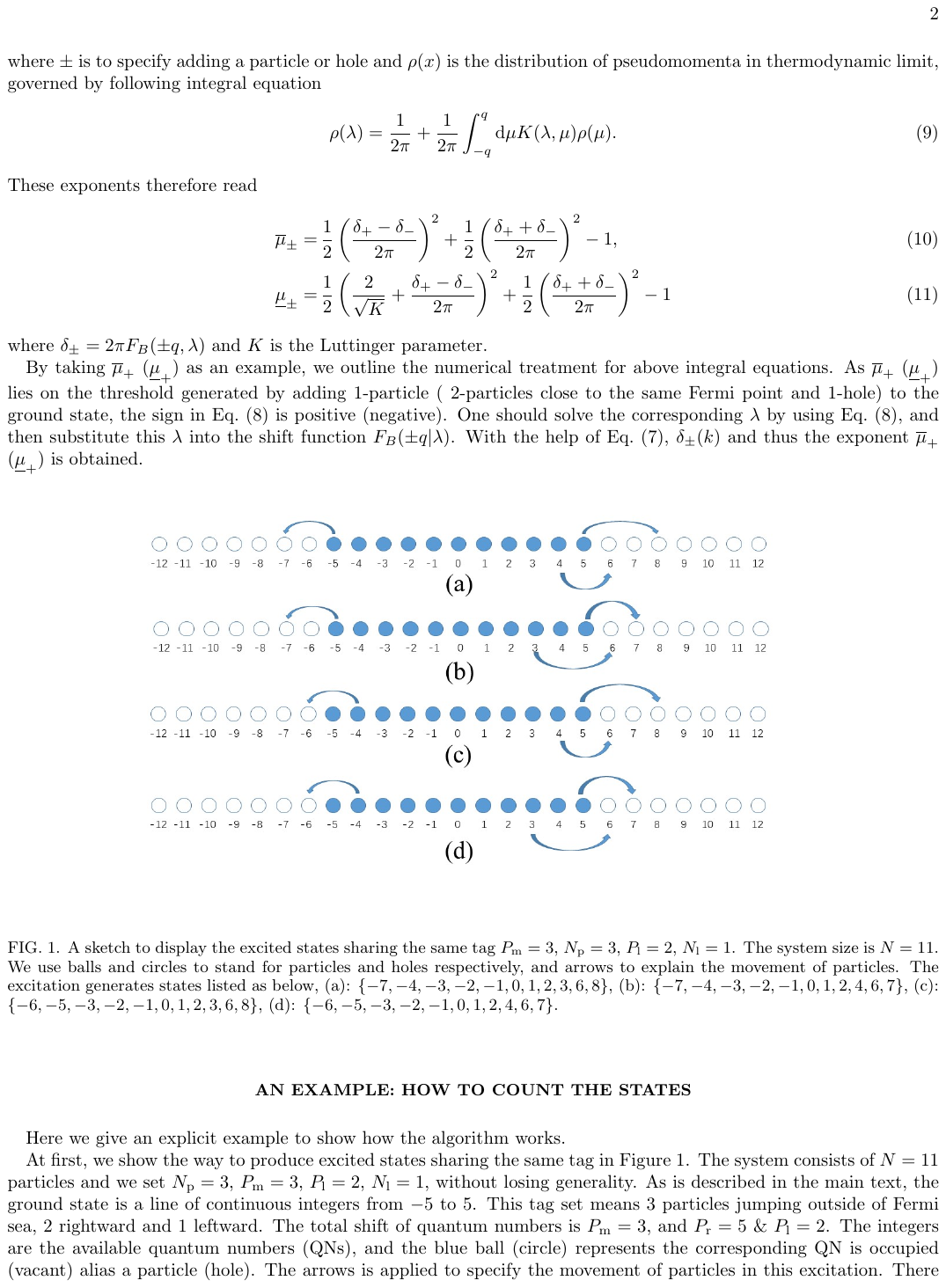}
\end{figure*}

\begin{figure*}[ht]
 \centering
 \includegraphics[width=\textwidth]{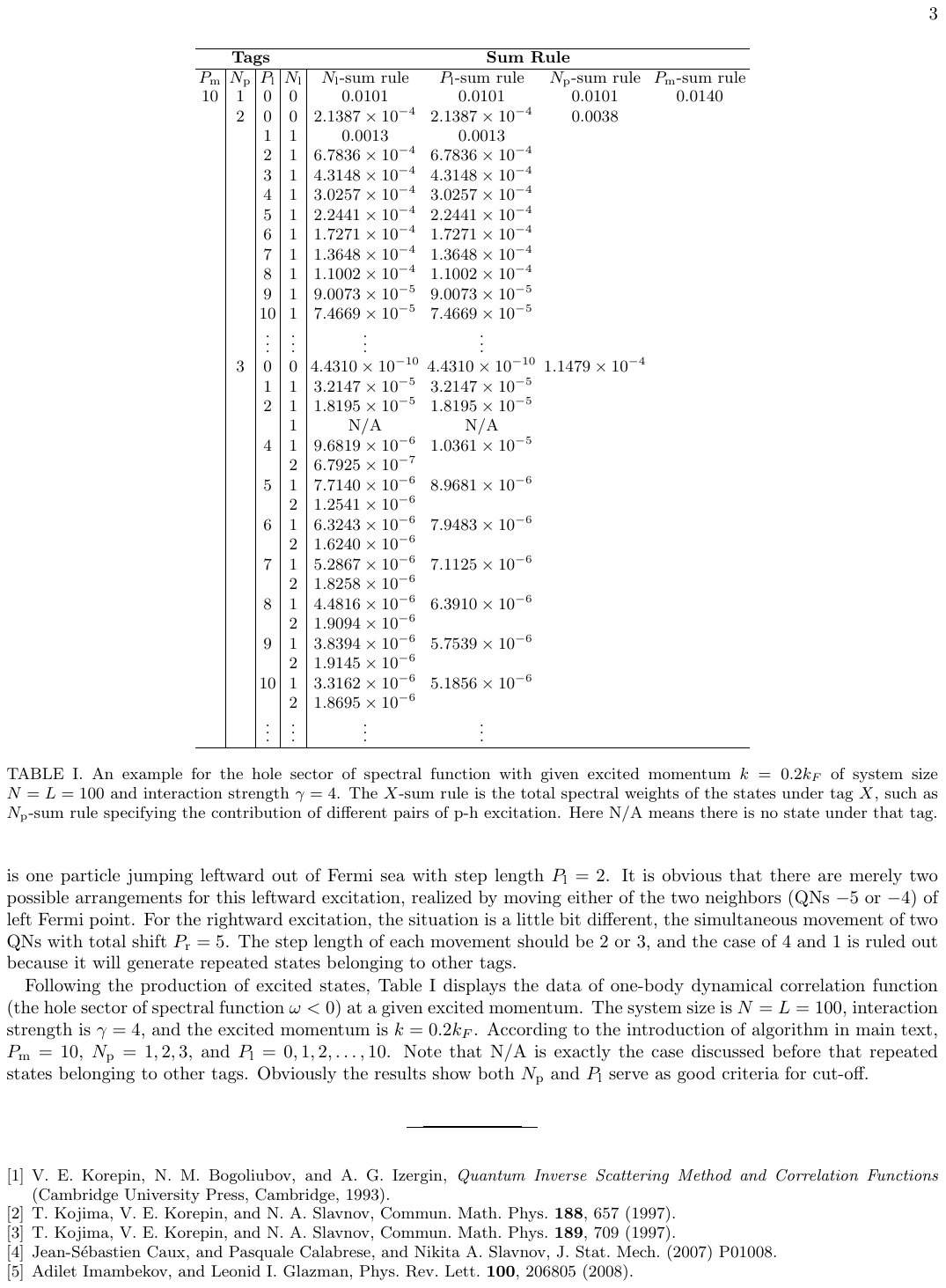}
\end{figure*}

\end{document}